\documentstyle[12pt]{article}

\begin{document}
\newcommand {\be}{\begin{equation}}
\newcommand {\ee}{\end{equation}}
\newcommand {\bea}{\begin{array}}
\newcommand {\cl}{\centerline}
\newcommand {\eea}{\end{array}}
\def\o{\over}
\def\par{\partial}
\baselineskip 0.65 cm
\begin{flushright}
IPM-97-216 \\
hep-th/9709054
\end{flushright}
\begin{center}
\bf \Large {Mixed Boundary Conditions and Brane-String Bound States}
\end{center}
\cl {\it H. Arfaei}
\cl {\it and}
\cl{\it M.M. Sheikh Jabbari}
\cl {\it Institute for studies in theoretical Physics and mathematics 
IPM} \cl{\it P.O.Box 19395-5746, Tehran,Iran}
\cl{\it Department of physics Sharif University of technology}
\cl{\it P.O.Box 11365-9161}
\cl{\it e-mail  Arfaei@ theory.ipm.ac.ir}
\cl{\it e-mail  jabbari@netware2.ipm.ac.ir}
\vskip 2cm
\begin{abstract}
 In this article we consider open strings with mixed boundary conditions
(a combination of Neumann and Dirichlet conditions at each end). We discuss 
how their end points show a $D_p$-brane with NS-NS charge, 
i.e. {\bf a bound state} of a D-brane with a fundamental strings. We show that
these branes are BPS saturated. In the case of one-branes, we show that
their mass densities are in agreement with IIb SUGRA which is
Sl(2,Z) invariant. Using Chan-Paton factors, we extend our results to 
the case of bound states of $n$ D-strings and $m$ F-strings. 
These string theoretic results are also checked in the effective field 
theory limit.

\end{abstract}
\newpage
\section {Introduction}
 
 Since their discovery by Polchinski [1,2], D-branes have contributed immensely 
to our understanding of various phenomena in superstring theories.
By considering their interactions via the exchange of closed strings, one can
recover their long range behaviour which get contributions from massless 
fields graviton, dilaton, and RR fields.
One finds that they are BPS saturated objects with a unit of the proper 
RR charge and that their mass density is proportional to ${1 \o g_s}$ [1], 
as predicted by the effective field theories (SUGRA). It has also been shown that 
we can construct bound states of such objects with themselves [3,4,5] or 
with F(undamental)-strings [3]. These bound states may or may not be BPS states. 
The bound states can be constructed from branes of the same or different 
dimensions. Among these bound states of $p-$ and $p'$-branes, those with $p'=p+2$ 
are truely bound states[4], i.e, their energy is lower than the sum of 
the individual energies.
For $p'=p+4$ or $p'=p$, due to SUSY, they are marginally bound 
( thier masses are exactly the sum of the masses of the individual branes). 
In the case of bound state of $n$ similar D-branes, one can check that the 
corresponding brane system is BPS saturated carries $n$ units of RR charge
[3]. In this case, open strings attached to D-branes look like Chan-Paton 
gauged strings with gauge group $U(n)$.

 Besides bound state of D-branes, we can have BPS bound states of F-strings 
with D-branes.  Witten has shown that these states carry the charge of a $U(1)$
gauge field living in the D-brane [6,3], which is partly the pull back of the 
background Kalb-Ramond field. 
Since the $U(1)$ charge in D-string case is the momentum conjugate 
to $U(1)$ gauge field, the corresponding charge aquires integer values 
through quantization [3]. If we consider other $D_p$-branes, the electric $U(1)$ charge 
is again related to momentum and field strength of $U(1)$ field.

We show that the (F-string)-(D-brane) BPS {\it bound states} can be represented
in string theory by {\it mixed boundary conditions} 
on open strings attached to branes. The mixed boundary conditions has been first 
considered in [6]. A particular class of mixed boundary conditions have been 
perviously considered [5,7] to describe bound states of $p$ and $p-2$-branes, which do not carry 
NS-NS charge [8]. 
They are arrived at by applying T-duality to a direction which is niether 
perpendicular nor parallel to a $D_{(p-1)}$-brane. The boundary conditions are then 
those of a $D_p$-brane with an internal magnetic gauge field, i.e. the mixed boundary
conditions do not involve the time. Although these works considered mixed boundary 
conditions, their calculations are all within the context of SUGRA and not within 
string theory. Mixed boundary conditions have also been encountered in [9], 
where again magnetic Wilson lines are assumed to exist in the six dimensional
compact manifold of type I $(N=1)$ theories.

In this article the mixed boundary conditions are extended to include  branes
with internal electric field. As a result the mixed boundary conditions also 
involve the time direction. This extension will have novel implications:
the brane will become a bound state of a D-brane and a number of F-strings.
In contrast to the magnetic case [5,7], the branes we consider carry charge of
NS-NS Kalb-Ramond field. 
This fact can be observed in variety of ways. First, we can
derive it applying a chain of T-dualities, discussed in section 3. 
Second, considering the DBI action and the $B$ field equation of motion, we find that
$F_{0i}$ is proportional to the NS-NS charge density of bound state.
Third, within string theory, finding the long range interaction of such bound 
states, we show the existence of the $B_{\mu\nu}$ charge. 
The third argument constitues a generalization of the Polchinski's work to 
(F-string)-(D-brane) bound states. 

In general, we will use the $(m,\ 1_p)$ notation for bound states of $D_p$-branes
and F-strings, in which $m$ determines the magnitude of NS-NS two form 
charge \footnote{The $B_{\mu\nu}$ charge is a two form defined in the world-volume 
of brane.}, and $1_p$ determines charge of the proper RR form ((p+1) RR form).

As discussed in the sequel, the mixed boundary method is capable of representing 
the NS-NS charge \footnote{ Unless it is mentioned explicitly, by NS-NS charge we
mean the NS-NS two form charge.}, i.e. ,they enable us to construct a string 
theoretic description of these bound states.

 Here we mainly focus on one-branes in IIB theory. These objects
are needed for Sl(2,Z) duality of IIB strings [10]. Using the conventional 
notation we denote their charges (i.e.,  NS-NS, Kalb-Ramond, and RR two form, respectively) 
charges by $(m,\ n)$, an Sl(2,Z) doublet. 

In section 2, we review the derivation of mixed boundary conditions from 
a $\sigma$-model action [6] and by means
of open strings stretched between two $(m,\ 1)$ strings. Using the standard 
techniques of [1], we calculate their interactions. Their vanishing  
indicates that the branes defined by mixed boundary conditions are BPS saturated 
and preserve one-half of the SUSY, like (0, 1) strings introduced by Polchinski 
[1]. 
 We also obtain their mass density which again is a check for BPS conditions. 
The problem of these bound states in the case of $P$-branes 
($(m,\ 1_p)$ branes ,$p>1$) is also discussed. 

In section 3, we will analyze how NS-NS charged branes of IIB theory 
are related to moving $D$-branes in IIA string theory under T-duality.
Our discussion sheds light on the problem of the relation between Sl(2,Z)  
symmetry of IIB string theory and T-duality of IIA and IIB theories [11].

In section 4, we discuss how $(m,\ 1_{p})$ bound states, similar to $(0,\ 1_p)$ 
states, show symmetry enhancement when they are 
coincident. Moreover, we explain how these symmetry enhancements in the case of
IIB D-strings or $D_5$-branes also hold in the strong coupling regime. 
Equivalently, we will present a Sl(2,Z) invariant argument for symmetry 
enhancement. 

In section 5, we show how $(m,\ n_p)$ branes can be described by means of both 
Chan-Paton gauged open strings and by mixed boundary conditions. Calculating the
interactions of two $(m,\ n_p)$ branes, we obtain their mass density and discuss their symmetry enhancements 
in the coincident limit.

In section 6, we present the field theoretic description of what we have done
in previous sections. We do this using the generalized ( Sl(2,Z) invariant) DBI-type 
action [12] and IIb SUGRA [5]. Again we calculate the $(m,\ n)$ string interactions 
at the tree level from field theory which is in agreement with the long range 
results of string theory. 

In section 7, we will discuss some new features and open questions.

\section {Mixed Boundary Conditions:}

 One can introduce Dirichlet boundary conditions on open strings by 
adding a constraint term to usual $\sigma$-model [6] as following:

\be 
S= {1 \over 4\pi\alpha'} \int_{\Sigma} d^2\sigma \bigl[ G_{\mu\nu}\partial_aX^{\mu}
\partial_bX^{\nu}g^{ab}+ \epsilon^{ab} B_{\mu\nu}\partial_aX^{\mu}\partial_bX^{\nu}     
+ \alpha' \partial^a \rho \partial_aX^{\mu}\partial_{\mu}\Phi \bigr]+\\
{1 \over 2\pi\alpha'}\oint_{\partial \Sigma} d \tau A_i \partial_{\tau}\zeta^i, 
\ee
where $\Sigma$ and $\partial\Sigma$ are the world-sheet and its boundary
, $A_i$ $i=0,1,...,p,$  the $U(1)$ gauge field living in a $D_p$-brane,
$\zeta^i$ are its internal coordinates and $G,B,\Phi$ are usual back-ground 
fields. Variation of this action with respect to $X^{\mu}$ gives either of the 
following boundary conditions:

\be
\left\{  \begin{array}{cc}
\delta X^{\mu}=0 \\ 
G_{\mu\nu}\partial_{\sigma}X^{\nu}+F_{\mu\nu}\partial_{\tau} X^{\nu}=0  
\end{array}\right.
\ee
where
\be
F_{\mu\nu}=\left\{  \begin{array}{cc}
B_{\mu\nu} \;\;\;\;\;\ \mu >p \\  
B_{\mu\nu}-A_{[\mu,\nu]} \;\;\;\;\;\ \mu,\nu \leq p.
\end{array}\right.
\ee
 
  As we see when there is a non-trivial $U(1)$ gauge field strength, 
$F_{\mu\nu}$, the usual Neumann boundary conditions is replaced by mixed 
boundary condition. So it suggests that we are able to introduce bound state 
of D-branes with F-strings (carrying a non-vanishing $B_{\mu\nu}$ field with
a source on the brane and 
hence making a non-trivial $F_{0\nu}$ background) by mixed boundary condition 
on open strings attached to.
A similar boundary conditions with non-zero $B_{\mu\nu}$ ($\mu,\nu\neq 0$)
have been considered to discuss bound state of $p, p-2$ branes [5] in the 
context of SUGRA theories. It is worth noting that in those cases the $F_{\mu\nu}$  
shows the distribution of $p-2$ on the $p$-brane world-volume. 
Similarly in our case ($F_{0i}\neq 0$), $F_{0i}$ for the ($p>1$)
shows the NS-NS charge distribution. The $B_{\mu\nu}$
field found in [5] are source free unlike what we obtain here.
In this article we mainly focus on D-string case. 

For $(m,\ 1)$ string, following [3] we can argue that $F_{ab}=m \lambda \epsilon_{ab} ,
\;\ a,b=0,1$, where 
$\lambda$ is the string coupling constant. So the related boundary conditions 
are: 

\be
\left\{  \begin{array}{cc}
\delta X^{\mu}=0 \;\;\;\;\ \mu=2,...,9 \\ 
\partial_{\sigma}X^a+m\lambda\epsilon_{ab}\partial_{\tau}X^b=0.
\end{array}\right.
\ee
the corresponding fermionic boundaries are

\be
\left\{  \begin{array}{cc}
(\psi^{\mu}-(\frac{im-1}{im+1})\tilde{\psi}^{\mu})|B>=0 \;\;\;\ \mu=0,1\\
(\psi^{\mu}+\tilde{\psi}^{\mu})|B>=0 \;\;\;\ \mu=2,..,9.
\end{array}\right.
\ee

The value of the quantum of charge can be determined from the calculation
on the disk, or equivalently by use of a one-loop vacuum amplitude. To do so  
let us consider two parallel $(m,\ 1)$ strings at $X^{\mu}=0$ and $X^{\mu}=Y^{\mu}$
$\mu=2,..,9$. The one loop vacuum graph of mixed open strings (open strings
with mixed boundary conditions) is sum over the cylinders with ends on each 
D-brane. In the closed strings channel this is exchange of single closed string
between them.

Before the mode expansion for the mixed open strings stretching between branes and
build their first quantized components.

Imposing (4) on $X^{\mu}(\sigma,\tau)$ at $\sigma=0,\pi$ we have:

\be
\left\{  \begin{array}{cc}
X^{0}=x^0+N(p^0 \tau-m\lambda p^1 \sigma)+N' \sum_{n\neq 0} {e^{-in\tau} \over n}
\bigl(a^0_n \cos n\sigma + im\lambda a^1_n \sin n\sigma \bigr) \\
X^1=x^1+N(p^1 \tau+m\lambda p^0 \sigma)+N' \sum_{n\neq 0} {e^{-in\tau} \over n}
\bigl(a^1_n \cos n\sigma - im\lambda a^0_n \sin n\sigma \bigr) \\
X^{\mu}=Y^{\mu} {\sigma \over \pi} +\sum_{n \neq 0} a_n^{\mu} {e^{-in\tau} \over n}
\sin n\sigma \;\;\;\;\ \mu=2,..,9,
\end{array}\right.
\ee
where $N, N'$ are some normalization factors which are determined by considering 
the canonical commutation relations
\be
[X^0,P^0]=[X^1,P^1]=i.
\ee

As it is seen from the action (1) the conjugate momentum of the $X^0,X^1$ 
are:

\be
\left\{ \begin{array}{cc} 
P^0=\partial_{\tau}X^{0}+F_{01} \partial_{\sigma} X^{1} \\ 
P^1=\partial_{\tau}X^{1}-F_{10} \partial_{\sigma} X^{0}  
\end{array}\right.
\ee
 
By a gauge transformation introduced in [3] which is only a function of 
$(m,\ 1)$ string coordinates, $F_{ab}$ can be set to be 
$m\lambda\epsilon_{ab}\delta^{\mu}_a\delta^{\nu}_b$. It is worth to note 
that the corresponding gauge transformation
do not change the boundary conditions which is only a function of gauge 
invariant F.

For general $F_{\mu\nu}$ the extra term in the canonical momentum densities  
looks like world sheet gauge potentials.
For constant $F$ it only leads to a change in the normalization factor $N$
which is obtained to be:

\be 
N={1 \over 1+m^2\lambda^2} ,\;\;\;\;\;\ N'=N^{1/2}.
\ee
They lead to the expected commutation relations:

\be 
[a_{n}^{\mu},a_{m}^{\nu}]=\delta_{n+m} \delta^{\mu \nu} \;\;\;\ \mu,\nu=0,..,9.
\ee
The amplitude for the exchange of closed string is:

\be 
A=\int {dt \over 2t} \sum_{i,p}e^{-2\pi\alpha' t \cal H },
\ee 
where $i$ indicates the modes of the open string and $p$ the momentum which has 
non-zero value in 0,1 components,
and $\cal H$ is the open string world-sheet Hamiltonian, which is
obtained from action (1). Performing integration
on momentum part and trace on oscillatory modes we get:

\be
A=2V_2 (1+m^2\lambda^2) \int \frac{dt}{2t}(8\pi^2\alpha't)^{-1}e^{-\frac{Y^2t}{2\pi^2\alpha'}} 
 (\bf{ NS-R}), 
\ee
where $\bf{ NS}$ and $\bf{ R}$ are given by 
\be
{\bf NS} = {1 \over 2}q^{-1}\prod{\biggl(\frac{1+q^{2n+1}}{1-q^{2n}}\biggr)}^8
-{1 \over 2}q^{-1}\prod{\biggl(\frac{1+q^{2n-1}}{1-q^{2n}}\biggr)}^8 .\\
\ee
\be
{\bf R}= 8\prod{\biggl(\frac{1+q^{2n}}{1-q^{2n}}\biggr)}^{8}.
\ee
The novel feature of this result is the multiplicative factor involving $\lambda$
which modifies the tension.
The Total interaction vanishes, which is a sign of SUSY in open string channel
(the corresponding solutions preserve half of the SUSY). In the exchanged closed string 
point of view this is a sign of BPS condition for the branes [13]. We can extract
graviton and dilaton and RR, Kalb-Ramnod contributions. In order to see the effective
low energy contributions (massless closed strings) we go to $t\rightarrow 0$
limit:

\be
A={\bf[1-1]}(1+m^2\lambda^2)V_2 (2\pi)(4\pi^2\alpha')^2 G_8(Y^2).
\ee

The ${\bf -1}$ term in brackets is due to graviton, dilaton and the ${\bf 1}$
term is due to the contribution of the RR and NS-NS two forms respectively; 
i.e. the term proportional to $m^2\lambda^2$ is due to $B_{\mu\nu}$ and $1$ 
due to RR two form.

Vanishing amplitude as we will see explicitly in field theoretical calculations 
shows the BPS condition saturation and reminds the "No Force Condition" between  
BPS states which is also a sign of SUSY.

 From now on we use the Sl(2,Z) doublets consisting of the RR and NS-NS 
two forms,  ($\tilde{B}_{\mu\nu}$) and $B_{\mu\nu}$: 
\be
{ \cal B}_{\mu\nu}=\left( \bea{cc}
B_{\mu\nu} \\
\tilde{B}_{\mu\nu}
\eea\right). 
\ee

Comparing with field theoretical rsults, we show explicitly in section 6 that 
how from (15) we get the $(m,\ 1)$ string BPS mass density formula (when the RR 
scalar vanishes) to be:

\be
T_{(m,1)}=(4\pi^2\alpha')^{1/2}\sqrt{{1 \over \lambda^2}+ m^2} . 
\ee
where  $(4\pi^2\alpha')^{1/2}$ is the F-string tension. 
 The above results can easily be generalized to some special
$P$-brane cases which are described by only non-zero $F_{0i}$.
In these cases by a coordinate  transformation in world
volume we can obtain a $F_{0i}$ with only two non-zero components, e.g. the 
$F_{01}$ and $F_{10}$.

In general $B_{\mu\nu}$ charge is a two form  defined on the
D-brane world-volume which in the case of string is just proportional to 
$\epsilon_{ab}$ in two dimensional world-sheet.

In these special cases the calculations presented here for the $(m,\ 1)$
strings, is not much altered. For $(m,\ 1_p)$ branes only two components of 
attached open strings have mixed, $(p-1)$ of them Neumann 
and the transverse $(9-p)$ components by Dirichlet boundary conditions.
It is crucial that the mixed boundary conditions is imposed on $X^0$, and
an arbitrary $X^i$. Thus introducing a non-zero $F_{0i}$ field, breaks the
$SO(p,1)\times SO(9-p)$ lorentz symmetry to $SO(1,1)\times SO(p-1)\times 
SO(9-p)$. This anisotropy gives rise to a new intrinsic form i.e, the volume  
form in the $(0,i)$ subspace. The bulk fields can also be along this new form, 
in contrast to D-branes where only the total volume form is consistent with the
symmetries which allows only non-vanishing $(p+1)$ forms. 

 For the case of two {\it similar} $(m,\ 1_p)$ branes interaction, again we 
obtain the result for the interaction of two similar $D_p$ brane multiplied by 
the factor $(1+m^2\lambda^2)$.
Again vanishing of the amplitude signals presence of super symmetry and BPS
property of the $(m,\ 1_p)$-brane bound states.
More over the tension; the constant in front of the amplitude, is 
\be
T_{(m,\ 1_p)}=T^0_{(p)} \sqrt{1 + \lambda^2 m^2}
\ee
where $T^0_p$ is tension of a $D_p$-brane.

The state $(m,1_p)$ inspite of being anisotropic is uniform which can be seen  
from the amplitude being proportional to the Green's function in the transverse 
space and also from the boundary conditions. Therefore the strings are uniformly 
distributed on a P-brane world-volume with density proportional to  $m$.
This explains the finite contribution of F-strings to the tension of   
bound state. In the world-volume super symmetry point of view, this is 
the share of the NS-NS charge density to the tension of the BPS state.

The system of two $(m,\ 1_p)$ branes with non-parallel NS-NS  charges; their interactions
does not vanish any more (the graviton,dilaton contrubutions do not 
cancel RR and Kalb-Ramond's). Interaction of these branes is studied
in field theory limit, which we will explain it in section 6. 

\section {T-duality of IIA , IIB and $(m,\ 1_p)$ Bound states:}  
 
 It is well known that, e.g.[14], the type IIA theory on a circle is T-dual 
to IIB. The problem we would like to address in this section is the identification 
of the states corresponding to $(m,\ 1_p)$ branes after T-duality.
Let us consider a $D_{p-1}$-brane defined by the following boundary conditions
for the end of the strings attached to it [15]:

\be
\left\{\bea{cc}
\delta X^{\mu}=0 \;\;\;\;\ \mu=p+1,...,9 \\ 
\partial_{\tau}(X^{p}- \Omega^p_{\alpha} X^{\alpha})=0  \\
\partial_{\sigma}(X^{\alpha}+ \Omega^{\alpha}_p X^{p})=0  \\
\partial_{\sigma} X^{\mu}=0 \;\;\;\;\ \mu=0,..p-1, \mu\neq \alpha.
\end{array}\right.
\ee

If $\alpha=0$, these boundary conditions show a moving $D_{p-1}$-brane in the 
P-direction with velocity $\Omega_{0p}$, and if $\alpha\neq 0$ they represent a 
$D_{p-1}$-brane rotated in $(\alpha, p)$ plane with angle $\theta=tg^{-1}
(\Omega_{\alpha\ p})$.
Under T-duality in $X^p$ direction, $\partial_{\sigma}$ and $\partial_{\tau}$ 
acting on $X^p$, are interchanged. Hence (19) is replaced with: 
\be
\left\{\bea{cc}
\delta X^{\mu}=0 \;\;\;\;\ \mu=p+1,...,9 \\ 
\partial_{\sigma}X^{p}- \Omega^p_{\alpha} \partial_{\tau}X^{\alpha}=0  \\
\partial_{\sigma}X^{\alpha}+ \Omega^{\alpha}_p \partial_{\tau}X^{p}=0  \\
\partial_{\sigma} X^{\mu}=0 \;\;\;\;\ \mu=0,..p-1, \mu\neq \alpha.
\end{array}\right.
\ee

describing a $(m,\ 1_p)$ brane for the $\alpha=0$ with
\be
\Omega_{0p}=F_{0p},
\ee

and for $\alpha\neq 0$ a $(p, p-2)$ brane bound state [5] in which $p-2$ branes
are homogenuously distributed on $P$-brane world volume in $(\alpha,p)$ plane
with:
\be
\Omega_{\alpha p}=F_{\alpha p}.
\ee
It can be seen that  in the case of our interest, the internal electric
gauge field ($F_{0i}$), has aquired non-zero value under T-duality in the 
velocity direction. This means that in T-dualizing a type II theory(A or B),
a $(p-1)$-brane with momentum\footnote{R is the radius of compactification 
in the $X^p$ direction.},$m/R\ $ is transformed to a bound state of a $D_p$-brane and
$m$ F-strings.
The NS-NS charge density of this bound state is $m\lambda$.
$\lambda$ appears because of the relation between velocity and momentum, 
involving $(p-1)$-brane mass density: $\;\;\ {\alpha'^{p/2} \over \lambda}$. 

A particular result of the above is that, a dual state corresponding to NS-NS 
charged strings is the momentum modes of the $D_0$-branes.

We have shown that one can get $(m,\ 1_{p+1})$ brane of IIB (IIA) theory by
T-dualizing moving $D_p$-branes of type IIA (IIB). Therefore $(m,\ 1)$ strings
with different NS-NS charges are related by a boost transformation in the 
T-dual theory.
\vskip 0.5cm

\centerline{$ (m,\  1)string \hspace{4.5cm} (0,\ 1)string$ } 

\hspace{5cm}  $\mid$  \hspace{5cm} $\mid$ 

\hspace{3.1cm} T-duality $\uparrow$  \hspace{4.95cm} $\uparrow$ T-duality

\hspace{5cm}  $\mid$  \hspace{5cm} $\mid$ 

\centerline{ $(v=m\lambda)\ D_0-brane\stackrel{BOOST}{---\rightarrow---} (v=0)\ D_0-brane$ }

\vskip 1cm

In the above diagram T-duality is done in boost direction.

It may seem that a Sl(2,Z) transformation on type IIB [10] may project $(m,\ 1)$
state to $(m,\ 1)$ and hence complete the above diagram. Specifically the $T$
subgroup of Sl(2,Z) maps $(0,\ 1)$ to $(m,\ 1)$ but the obstacle is that such
transformations also moves the type IIB theory in its moduli space and transforms
the RR scalar from zero to $m$. On the other hand T-duality on type IIA theory
takes it to a type IIB thoery with zero $\chi$ (the RR scalar). 
To go beynd $\chi=0$ we need to consider the M-theory compactified on $T^2$
instead of ${IIA  \over S^1}\ \ ((M/S^1)/S^1)$.

\section {Bound State of $(m,\ 1_p)$ branes and Symmetry \newline 
Enhancement:}

Similar to the coincidence limit of $D_p$-branes [3], corresponding limit of 
$(m,\ 1_p)$ branes also shows symmetry enhancement. When $n$ $(m,\ 1_p)$ branes 
coincide their internal field theory becomes a $U(n)$ SYM theory in (1+1) dimensions. 

There are $n$ massless string states with their two ends on each of them. 
Between any two branes, $i$ and $j$, two oriented strings can be suspended 
with mass proportional to \newline
$|Y_i-Yj|\sqrt{1+m^2\lambda^2}$ which vanishes in the
desired limit; $Y_i=0\;\,i=0,...n$. In this limit we have a total number of 
$n^2$ massless states forming the gauge vector for a $U(n)$ symmetry.
To clarify the above argument for $(m,\ 1)$ string let us study the internal
field theory of it. 

In the case of $D_p$-brane or $(0,\ 1_p)$, the gauge field strength is partly 
pull back of $B_{\mu\nu}$:
\be
F_{ab}=B_{ab}-A_{[a,b]}.
\ee
In the case of  $(m,\ 1)$ string of type IIB Sl(2,Z) duality requires the existence  
of an extra $\tilde{A_a}$ field which is the pull back of $\tilde{B}_{\mu\nu}$, the RR
two form [12].
The field strength of it is given by
\be
\tilde{F_{ab}}=\tilde{B_{ab}}-\tilde{A_{[a,b]}}.
\ee
It is clear that $F$ and $\tilde{F}$ form a Sl(2,Z) doublet
\newline${ \cal F}_{\mu\nu}=\left( \bea{cc}
F_{\mu\nu} \\
\tilde{F}_{\mu\nu}
\eea\right).$ 

It may seem that we have two $U(1)$ gauge fields living in the $(m,\ 1)$ string
world-sheet. This is not the whole story; a particular combination of  the two 
gauge fields $F$ and $\tilde{F}$ decouples from the $(m,\ 1)$ string in the 
classical level.
This leaves us with only one $U(1)$ gauge field in the $(m,\ 1)$ string 
world-sheet. To find this particular combination we use the Sl(2,Z) invariant 
action for $(m,\ n)$ strings [11]. In this action which we return to it more 
specifically in section 6, the classical solution of equation of motion contains 
fluctuations of ${\cal F}$ which is perpendicular to state of $(m,\ 1)$ string 
in the moduli space. So among different ${\cal F}$ only the combination which 
satisfies this condition remains:
\be
(m,\ 1){\cal M}{\cal F}=0,
\ee
where ${\cal M}$ is the moduli space metric (33).
This equation is manifestly Sl(2,Z) invariant, hence we can find the particular
$U(1)$ gauge field from it: 

\be
{ \cal F}_{\mu\nu}=\left( \bea{cc}
F_{\mu\nu} \\
-m\lambda^2{F}_{\mu\nu}
\eea\right). 
\ee

So in the coincidence limit, i.e. when $n$ $(m,\ 1)$ strings are on top of each 
other they make a $(nm,\ n)$ string in which, there is a $U(n)$ gauge theory.

It is necessary to bear in mind that symmetry enhancement of coincident similar
$D_p$-branes for even $p$ is only valid for the usual string theory limits($g_s <<1$), at 
strong coupling for IIA cases (even p) symmetry enhancement arguments fails [16].
In the case of type IIB theory symmetry enhancement argument 
is supported by Sl(2,Z) duality and holds even in strong coupling.

\section  {$(m,\ n_p)$ brane bound states in string theory}

In this section we construct bound state of $m$ F- strings with $n$ $D_p$-branes
in string theory. 
For this purpose we first consider $n$ similar $D_p$-branes bound state,
and then by imposing mixed boundary conditions on two components ($X^0$ and $X^i$)
of open strings attached to brane, we obtain $n$ $D_p$-brane $m$ F-string bound
state. By means of this definition, we study $(m,\ n)$ strings interactions 
in string theory limit ($g_s <<1,\ \chi$=0). As a result we find
their mass or charge density saturating the BPS condition, in agreement with
SUGRA [10].

The open strings attached to $n$  $D_p$-brane system form an adjoint 
representation of U(n) gauge field [3].

The group theoretic state of these open strings is easily
introduced by usual Chan-Paton factors given to each end. These Chan-Paton
factors are in fundamental representation of U(n) (quark and anti-quark), 
so that the whole open string sits in adjoint representation of U(n). 
Explicitly one can write boundary condition of such open strings as:
\be 
\left\{  \begin{array}{cc} 
X^{\mu}=0 \;\;\;\;\ \mu=p+1,...,9 \\ 
\partial_{\sigma}X^{\mu}=0 \;\;\;\;\ \mu=0,...p
\end{array}\right.
\ee
where $X^{\mu}$ is an $n\times n $ $U(n)$ matrix. At each end ($\sigma=0,\pi$) we  
represent its U(n) state by ($\lambda_i,\bar{\lambda}_j$) or ($\bar{\lambda_i},
\lambda_j$) which $\lambda$ ,$\bar{\lambda}$ show the quark anti-quark 
representations \footnote {these two distinct states show that these open strings
are oriented strings.}.

Following above in presence of two $n$ and $n'$ $D_p$-branes the $X^{\mu}$
become $U(n+n')$ matrices. The $U(n+n')$ matrix can be divided in an obvious
manner to four parts; an upper corner $U(n)$ part representing the states of 
strings ending at both ends on the $n\ D_p$-brane, the lower corner $U(n')$
which represents the state of strings at both ends attached to the $n'\ D_p$-brane 
and two $n\times n' $ and $n'\times n $ matrices for strings stretched between the two
branes. 
So in the calculation of the interaction between $n\ ,\ n'$ parallel $D_p$-branes 
the sum on all the open string states includes a trace over the group theoretic
states of the strings stretched between branes which are in the $U(n)\times U(n')$ 
representation of $U(n+n')$ ). 
The amplitude is simply the same amplitude for nteraction of two unit RR charged 
D-branes times $nn'$. Vanishing of the amplitude shows that $n$ $D_p$-branes
on top of each other form a marginal BPS saturated state with mass proportional to 
${n \over g_s}$.

More generally we can extend the mixed boundary conditions to this group theoretic
states. In this way we are able to construct bound state of $m$ F- strings
with any number of $D_p$-branes.

Here we restrict ourselves to $(m,\ n)$ strings whose boundary conditions are:

\be
\left\{  \begin{array}{cc}
\delta X^{\mu}=0 \;\;\;\;\ \mu=2,...,9 \\ 
n \partial_{\sigma}X^{a}+m \lambda \epsilon_{ab}\partial_{\tau} X^{b}=0  
\end{array}\right.
\ee

and $X^{\mu}\;\;\; {\mu=0,...,9}$ in adjoint representation of U(n). So
their state at $\sigma =0, \pi$ are given by $\lambda_i$ or $\bar{\lambda_j}$. 

By this method we can calculate the $(m,\ n)$ string interactions through the 
one-loop vacuum amplitude of these mixed Chan-Paton open strings:
\be 
A=n^2 (1+{m^2\lambda^2 \over n^2})\times A_0,
\ee
where $A_0$ is the corresponding amplitude for two (0 1) strings. Thus at 
massless closed strings or effective field theory limit one can write (29) as:
\be
A= (4\pi^2\alpha')({ n^2 \over \lambda^2}+m^2) V_2 (1-1) G_8(Y^2).
\ee

The above result shows that $(m,\ n)$ string (like $(m,\ 1)$ strings) form a 
BPS bound state with mass density $(4 \pi^2 \alpha')^{1/2}\sqrt{ m^2+ n^2/ 
\lambda^2} $. This is what one would expect from Sl(2,Z) symmetry.
We note that in general two $(m,\ n)$ and ($m',\ n'$) string case, the 
relative interactions do not cancel (except when ${m \over n}={ m'\over n'}$).

The gauge symmetry remaining after the formation of bound state (m,n) string
(considering mixed boundary conditions) is $U(r)$ where $r$ is the greatest
common divisor of $m$ and $n$. This is easily seen from a Sl(2,Z) transformation
which takes the $(m,n)$ to $(0,r)$.

Although the above arguments are given for strings, we can generalize it to 
any $(m,\ n_p)$ brane (bound state of 
$n$ $D_p$-branes with $m$ F-strings).
In this case the mass density of such 
bound state is obtained to be (18) 
multiplied by $n^2$ due to the trace 
on group theoretic states.
This shows that $(m,\ n_p)$ branes are 
BPS saturated bound states of
$D_p$-branes and F-strings. These bound state are related to (m,n) strings
by a chain of T-dualities.

In the case of IIB $(m,\ n)$ strings there is a Sl(2,Z) invariant combination  
of ${\cal F}$ components which is the surviving gauge field, like eqation (25).
The corresponding combination is determined from orthogonality condition:

\be
(m,\ n) {\cal M}\left( \bea{cc} 
F \\
\tilde{F}
\eea \right)=0
\ee

More precisely every string state in moduli space of IIB theory is given by a 
$(m,\ n)$ Sl(2,Z) doublet. The transverse oscillations of these strings
in the moduli space are described by $U(r)$ gauge fields related to massless 
states of attached open strings.

By a discussion similar to section 3, we can see that T-daulity,
in string spatial direction, transforms a $(m,\ n)$ string to a bound 
state of $n$ $D_0$-branes moving with velocity $m\lambda$ in the compact 
direction.

\section {Field Theory Descriptions, Manifest Sl(2,Z):} 

In the previous sections we built the string theoretic description of bound 
states of D-branes with D-branes or F-strings. We can check our
results in the IR limit with SUGRA results. In this section we restrict 
ourselves to one-branes of IIB theory. The field theory action consists of two 
parts;
the IIb SUGRA action [5] and a generalized Sl(2,Z) invariant DBI-type action
[12] including an interaction term. The first describes the dynamics of the 
bulk NS-NS and RR two form fields, and the latter the dynamics of 
the $(m,\ n)$ strings and their interactions with the bulk.

{\it IIb SUGRA action}

In order to build a IIb action, we use T-duality between IIa and IIb theories
on an arbitary direction , doing so we get to the following action [5]:

\be
\bea {cc}
S_{IIb}={1 \o 2k^2} \int d^{10}x \sqrt {-g}
\bigl[R+ {1\o 4}tr(\partial {\cal M} \partial {\cal M}^{-1})  
-{1 \o 12} {\cal H}^t {\cal M} {\cal H} 
-{1 \o 480} (dA^{(4)}-{\cal B}^tS{\cal H})^2\bigr]+\\
{1 \o 4k^2} \int A^{(4)}\wedge{\cal H}^tS{\cal H}. 
\eea
\ee
 where we have used the notation of [10]:
${\cal M}$, a 2$\times$2 Sl(2,R) matrix is the metric in the moduli space  
\be
{\cal M}=e^{\phi}\left( \bea{cc} 
|\lambda|^2\;\;\;\ \chi \\
\chi \;\;\;\;\;\ 1
\eea \right).
\ee
with $\lambda=\chi +ie^{-\phi}$ ($\chi $and $\phi$ are RR and NS-NS scalars 
respectively). ${\cal B}_{ab}$ is the Sl(2,R) doublet (16) and ${\cal H}=d{\cal B}$.
$A^{(4)}$ is the usual self dual 4-form of IIb theory which is a Sl(2,R) singlet.
S is a constant $2\times 2$ Sl(2,R) matrix:
\be
S=\left(  \bea {cc}
  0 \;\;\;\ 1  \\
  -1 \;\;\;\ 0
\eea \right).
\ee
Although we are concerned with self dual field $A^{(4)}$ [11], one can impose 
self duality conditions on by hand
\footnote {As explained in [5] the coefficient of $dA^{(4)}$ is ${1 \o 480}$
for self duality.}. 

The manifest Sl(2,R) invariance of (32), is broken to Sl(2,Z) by quantum 
considerations.

{\it  Manifestly Sl(2,Z) Invariant DBI-type Action for $(m,\ n)$ Strings}

As recently has been discussed by Townsend [12], there are two U(1) induced 
gauge fields related to ${\cal B}_{\mu\nu}$, which form an Sl(2,R) doublet:
\be
{\cal F}_{\mu\nu}={\cal B}_{\mu\nu}-\left( \bea {cc}

A_{[\mu,\nu]} \\
\tilde{A}_{[\mu,\nu]}
\eea \right).
\ee

Using this doublet Townsend generalizes the DBI action to [12]:
\be
S_{DBI}=\int d^2\sigma {1 \o 2v}[det g + {\cal F}^t {\cal M} {\cal F}].
\ee

 This action is manifestly Sl(2,R) invariant. In order to calculate two $(m,\ n)$
strings interactions, we also need a vertex term added to (36), As usual we take 
the minimal coupling as:
\be
S_{int}=\int d^2\sigma  {\cal J}_{ab}^t {\cal M} {\cal F}_{ab},
\ee
where ${\cal J}_{ab}$ is the current of $(m,\ n)$ string. In the case of static
$(m,\ n)$ strings:
\be
{\cal J}_{ab}=\epsilon_{ab} \left( \bea  {cc}
    m\\
    n
\eea \right).
\ee
 So the full action governing the $(m,\ n)$ dynamics is:
\be
S=S_{IIb}^{bulk} + S_{DBI} + S_{int}. 
\ee

The action given above is manifestly Sl(2,Z) invariant. 

Before going to brane interactions in detail, let us analyze the action (39).
If we solve the equation of motion for $A,\ \tilde{A}$ we have:

\be
{\cal F}_{ab}=\epsilon_{ab} \left( \bea  {cc}
    m\\
    n
\eea \right)+ {\cal G}_{ab}.
\ee

where $m,n$ are two constants and $ {\cal G}_{ab}$ is normal to $(m,\ n)$ in the 
moduli space. By quantization arguments the conjugate momentum of
$A,\ \tilde{A}$ ($m,n$ respectively) could only aquire integer values.
Inserting (40) into (36) and solving the equation of motion for $v$:
\be
S_{DBI}=\int T_{(m,\ n)} d^2\sigma \sqrt{det g} 
\ee

which describes a BPS string with tension:
\be
T^2_{(m,\ n)}=(m,\ n){\cal M} \left( \bea  {cc}
                          m\\
                          n
                        \eea \right)=(m+n\chi)^2+\frac{n^2}{\lambda^2}.
\ee
The same argument holds for the usual DBI action which describes the arbitrary 
$(m,\ 1_p)$ brane ($p>1$).
By taking the constant electric field solutions for $A_a$, we obtain:
\be
S_{DBI}=\int T_{(m,\ 1_p)} d^{(p+1)}\sigma \sqrt{det g},
\ee
where $T_{(m,\ 1_p)}$ is given by (18), and $m\lambda$ is the quantum value
of NS-NS charge or conjugate momentum of $A$ field living in the $D_p$-brane. 
Hence the usual DBI action or the action (36) describes the dynamics of objects 
with the given mass densities, i.e. D-brane, F-string bound states.

 In order to calculate $(m,\ n)$ strings interactions we use the usual methods [17].
We have to perform the Casimir energy calculations which in the first order
(tree diagram) is due to exchange of a single graviton and
${\cal B}_{\mu\nu}$ fields:

\be 
\varepsilon (Y) T=-2k^2_{10} \int d^{10}x \int d^{10}\tilde{x} \bigl[
T_{\mu\nu}\Delta^{\mu\nu,\rho\sigma}T_{\rho\sigma}-{\cal J}_{ab}^t{\cal M} 
\Delta{\cal J}_{ab} \bigr],
\ee
where first term in the brackets shows graviton dilaton contributions, and the
second, ${\cal B}_{\mu\nu}$ interactions. $\Delta^{\mu\nu,\rho\sigma}\ ,\ 
\Delta$ are the corresponding propagators and $T_{\mu\nu}\ ,\ {\cal J}_{ab}$
the related currents:
\be
T_{\mu\nu}={1 \o 2} T_{(m,\ n)}\delta(x_{\perp})\left\{ \bea{cc} 
\eta_{\mu\nu}\;\;\;\;\ \mu,\nu=0,1 \\
0 \;\;\;\ otherwise.
\eea \right.
\ee
where $x_{\perp}$ shows the coordinates normal to string. If we use a gauge
in which the dilaton contribution is absorbed in gravity part:
\be
\Delta^{\mu\nu,\rho\sigma}=(\eta_{\mu\rho}\eta_{\nu\sigma}-\eta_{\mu\sigma}
\eta_{\nu\rho})\Delta.
\ee
Putting them together, we obtain the Casimir energy: 
\be
\varepsilon (Y) =-2k^2_{10}V_{(1)}[ {\cal J}^t{\cal M}{\cal J}
- T^2_{(m,\ n)}] G_8(Y^2), 
\ee
where 
\be
{\cal J}^t{\cal M}{\cal J}=(m+n\chi)^2+\frac{n^2}{\lambda^2}.
\ee
When the BPS condition is saturated:
\be
{\cal J}^t{\cal M}{\cal J}= T^2_{(m,\ n)}.
\ee
And the IR limit of string theory is recovered. So the objects defined by 
mixed and Chan-Paton boundary conditions are {\bf string theoretic realization 
of $(m,\ n)$ string bound states}.
{\section {Discussion:}

We have explicitly consrtucted the bound states of F-strings and 
$D_p$-branes ($(m,\ 1_p)$ branes) in string theory simply imposing mixed 
boundary conditions on open strings having their ends on them. Moreover  
we can construct $(m,\ n_p)$ bound states by both mixed boundaries and Chan-Paton 
factors on the ends of open strings. These {\bf bound states} are BPS saturated
and their mass formula is given by (18). 

Another way of checking their long distance (effective field theory)
behaivour, is studying the scattering of an string off these objects which we 
have not considered here. Through these calculation we can also check their  
NS-NS charge explicitly. 

As we argued these bound states can also be understood by T-duality plus boost
in corresponding dual theory.

When we deal with $(m,\ n)$ strings, we should notice that each of these two 
(mixed boundary conditions and Chan-Paton factors) do not give an Sl(2,Z) 
invariant interaction sparately, but as we have shown explicitly a combination 
can give an Sl(2,Z) invariant $result$. 
Being more precise we are using methods and concepts from open F-string 
theory which are not Sl(2,Z) invariant, in other words 
we are doing our calculations in a special point of moduli 
space $(g_s <<1, \chi=0)$. Choosing this special point immediately breaks 
Sl(2,Z) in calculations but not in the final results.

 The symmetry enhancement arguments again holds when we have mixed open 
strings stretched between branes, 
this shows a way to make an Sl(2,Z) invariant symmetry enhancement argument 
for IIB one-branes. Such an argument does not hold for strong coulping limit of
IIA theory. In this paper although we mainly discussed one-branes the
results seems to hold true for IIB five-branes. 
For IIB three branes again the symmetry arguments is well matched to
Sl(2,Z) because their RR charge is an Sl(2,Z) invariant. Thus in the case of 
three branes only RR charge determines the gauge group (unlike the 
$(m,\ n)$ strings)\footnote{ We can also extend our method to $(m\ 1_3)$ branes 
although their
NS-NS charge is not Sl(2,Z) invariant since their RR charge is, the symmetry 
enhancement argument is valid under a Sl(2,Z) transformation.}.

The question of simultanuous presence of two strings with different 
NS-NS charges $(m,m')$ is not addressed here.
Applying the same method we introduced in this paper 
( mixed boundaries on coordinates parallel to D-brane), 
the spatial coordinates becomes $non-commutative$. 
The non-commutativity is controled by $(m-m')$ factor.
On the other hand we know that $(m,\ 1)$ and $(m',\ 1)$ strings system 
do not form a BPS saturated state, this is easily seen from the field 
theory calculations.

Through field theory calculations, energy of the system discussed above is: 
\be 
\varepsilon=\sqrt {{1 \o \lambda^2}+m^2}\sqrt {{1 \o \lambda^2}+m'^2}-
({1 \o \lambda^2}+mm').
\ee
which vanishes at $m=m'$. This result can be generalized for $(m,\ n)$,$(m',\ n')$
strings
as:
\be
\varepsilon=T_{(m, n)}T_{(m', n')}-T_{(m, n)}^{(m', n')}.
\ee
where $T_{(m, n)}$is the BPS mass formula:
\be
T^2_{m, n}=(m,\ n){\cal M}\left( \bea {cc}
m \\
n
\eea \right)\;\;\;\;;
T_{(m, n)}^{(m', n')}=(m'\ n'){\cal M}\left( \bea {cc}
m \\
n
\eea \right).
\ee
The interaction vanishes if ${m\o n}={m'\o n'}$. It seems that non-commutativity is
related to the fact that these strings do not form a BPS state.

There are some other solutions in related field theories that
have masses proportional to ${1\o g^2_s}$, the NS five branes.
So one expects that they can be introduced in string theory too, but till now 
although some special polarizations of them have been found in M(atrix)-theories 
[18], unlike D-branes, there is no string theoretic description of them. 
We can also think of the bound state of NS five branes with 
fundamental strings or D-branes. Type IIA NS five branes (or their 
possible bound states) can be studied via M-theory five branes. 
Bound state of such branes i.e. H-monopole with F-strings and their T-dual
i.e. KK-monopoles with F-strings winding charge has been recently investigated
, both in the context of string theory and M-theory [19,20]. 
In these dyonic states like what we have argued the NS two form charge of a
H-dyon is related to momentum modes of compactified F-string. Moreover
one can construct bound states of NS-branes with D-branes.
If we compactify the M-theory on $S^1$ on a direction
making an angle with one of the internal coordinates of $M_5$-brane,
we obtain a bound state of $D_4$-branes and NS five branes in IIA theory.

 NS five branes in IIB theory are quite different; by means of Sl(2,Z) symmetry 
they are related to $D_5$-branes. The symmetry enhancement argument for them
is exactly like usual $D_5$-branes, i.e. There is an Sl(2,Z) invariant $U(1)$
gauge field living in the bound state of $(1,\ n)$ five branes and the same holds 
for NS five brane. Hence we expect the symmetry  enhancement
in the coincident limit. 


{\bf Acknowledgements}

The authors would like to thank M. Alishahiha, F. Ardalan, F. Mansouri for 
their useful discussions. The authors would also like to thank S. Randjbar-Daemi 
and ICTP, Trieste Italy for hospitality during the period part of this work was performed.
We are grateful to  C. Vafa for his friutful comments.

\end{document}